\documentclass[prl,twocolumn,showpacs,preprintnumbers,amsmath,amssymb,citeautoscript]{revtex4-1}

\usepackage{graphicx}
\usepackage{epstopdf}
\usepackage{amssymb, amsfonts, amsmath}
\usepackage{dcolumn}
\usepackage{bm}
\usepackage{color}

\emergencystretch=\maxdimen
\newlength{\figwidth}
\begin{document}

\title{Full-Gap Superconductivity Robust against Disorder in Heavy-Fermion CeCu$_2$Si$_2$}

\author{T.\;Takenaka$^1$}
\author{Y.\;Mizukami$^1$}
\author{J.\,A.\;Wilcox$^2$}
\author{M.\;Konczykowski$^3$}
\author{S.\;Seiro$^{4,5}$}
\author{C.\;Geibel$^4$}
\author{Y.\;Tokiwa$^{6,7}$}
\author{Y.\;Kasahara$^6$}
\author{C.\;Putzke$^2$}
\author{Y.\;Matsuda$^6$}
\author{A.\;Carrington$^2$}
\author{T.\;Shibauchi$^1$}

\affiliation{$^1$Department of Advanced Materials Science, University of Tokyo, Kashiwa, Chiba 277-8561, Japan}
\affiliation{$^2$H.\,H.\,Wills Physics Laboratory, University of Bristol, Bristol BS8 1TL, England.}
\affiliation{$^3$Laboratoire des Solides Irradi{\'e}s, {\'E}cole Polytechnique, CNRS, CEA, Universit{\'e} Paris-Saclay,
	F-91128 Palaiseau, France}
\affiliation{$^4$Max Planck Institute for Chemical Physics of Solids, 01187 Dresden, Germany}
\affiliation{$^5$Institute for Solid State Physics, IFW-Dresden, Helmholtzstrasse 20, 01069 Dresden, Germany}
\affiliation{$^6$Department of Physics, Kyoto University, Kyoto 606-8502, Japan}
\affiliation{$^7$Center for Electronic Correlations and Magnetism, Institute of Physics, Augsburg University, 86159 Augsburg, Germany}

\date{\today}

\begin{abstract}{
A key aspect of unconventional pairing by the antiferromagnetic spin-fluctuation mechanism is that the superconducting energy gap must have opposite sign on different parts of the Fermi surface. Recent observations of non-nodal gap structure in the heavy-fermion superconductor CeCu$_2$Si$_2$ were then very surprising, given that this material has long been considered a prototypical example of a superconductor where the Cooper pairing is magnetically mediated. Here we present a study of the effect of controlled point defects, introduced by electron irradiation, on the temperature-dependent magnetic penetration depth $\lambda(T)$ in CeCu$_2$Si$_2$. We find that the fully-gapped state is robust against disorder, demonstrating that low-energy bound states, expected for sign-changing gap structures, are not induced by nonmagnetic impurities. This provides bulk evidence for $s_{++}$-wave superconductivity without sign reversal.
}\end{abstract}

\maketitle

Theories of unconventional superconductivity, where the pairing is not mediated by phonons, have been developed over the past decades to explain superconductivity in strongly correlated materials such as heavy fermions and high-$T_c$ cuprates. These theories have been challenged by recent and surprising results on the heavy-fermion superconductor CeCu$_2$Si$_2$ \cite{Yamashita} which was the first discovered heavy fermion superconductor \cite{Steglich}, and as such, the first candidate for an unconventional superconducting state.
The fact that superconductivity in CeCu$_2$Si$_2$ emerges near a quantum critical point of antiferromagnetic order has led to the almost universally held conclusion that its superconductivity is  unconventional with Cooper pairing mediated by spin fluctuations. The essence of this mechanism is that the momentum ($\bm{k}$) dependent repulsive interactions can effectively pair the electrons as long as the superconducting gap $\Delta(\bm{k})$ changes sign in $\bm{k}$-space. Depending on the structure of the Fermi surface and spin-fluctuations, this can lead to different sign-changing gap structures such as $d$-wave in cuprates or $s_{\pm}$-wave in iron pnictides. For CeCu$_2$Si$_2$, early experiments such as nuclear quadrupole resonance relaxation rate \cite{Ishida,Fujiwara} and specific heat \cite{Arndt} suggested $d$-wave superconductivity with line nodes in $\Delta(\bm{k})$. Inelastic neutron scattering measurements have shown an enhancement of magnetic spectral weight at around $E \sim 2 \Delta$, which has been interpreted as a spin resonance expected for a sign-changing  $\Delta(\bm{k})$ also consistent with $d$-wave symmetry \cite{Stockert,Eremin}.

In contrast to this, recent experiments which have combined specific heat \cite{Kittaka}, penetration depth, and thermal conductivity measured down to very low temperatures have shown that gap nodes do not exist at any point on the Fermi surface of CeCu$_2$Si$_2$ \cite{Yamashita}.   This nodeless structure might still be explained by a spin-fluctuation mechanism if the points in $\bm{k}$-space where the gap changes sign do not coincide with the Fermi surface sheets, as is the case for most iron-pnictide materials.  Specific calculations for CeCu$_2$Si$_2$ have shown that spin-fluctuations can lead to an $s_{\pm}$-type structure but the  closeness and corrugations of the Fermi-surface sheets means that  accidental nodes are unavoidable \cite{Ikeda}.  However, a nodeless $s_{\pm}$ state cannot be ruled out by this alone because the experimental Fermi surface of CeCu$_2$Si$_2$ has not been fully determined.  Hence, experiments to specifically probe for the presence or absence of a nodeless sign changing gap structure are needed.

One such experiment is the effect of impurity scattering on $T_c$. It was demonstrated that for CeCu$_2$Si$_2$, increasing impurity scattering leads only to a very weak reduction in $T_c$ \cite{Yamashita,Adrian} which would appear to be inconsistent with a sign-changing $\Delta(\bm{k})$.  However, given the quantitative nature of this argument further experimental confirmation is needed. Previous attempts at phase-sensitive measurements focused on the Josephson effect between CeCu$_2$Si$_2$ and Al \cite{Poppe,Sumiyama}. Although a finite Josephson current and conventional Fraunhofer pattern were observed, the results are not conclusive because the polycrystalline nature of the samples used complicates the interpretation. Moreover, in such measurements the possibility of a surface-induced $s$-wave component cannot be ruled out \cite{Sumiyama}. Thus bulk measurements that are sensitive to a possible sign change in $\Delta(\bm{k})$ of CeCu$_2$Si$_2$ are desirable.

Here we report on systematic measurements of the temperature dependence of the magnetic penetration depth $\lambda(T)$ in single crystals of CeCu$_2$Si$_2$, where impurity scattering has been introduced in a controlled way by electron irradiation. The irradiation creates point-like defects that act as nonmagnetic scatterers, which in the case of sign-changing order parameters should induce Andreev bound states at low energies. In the $s_{\pm}$ case, therefore, the fully-gapped state is expected to change to a gapless state with low-lying quasiparticle excitations \cite{Wang}. Indeed, in some iron-based superconductors a change in low-temperature $\Delta\lambda(T)=\lambda(T)-\lambda(0)$ from an exponential $T$ dependence to a $T^2$ dependence with increasing defect level is found which supports the model of an $s_{\pm}$ gap structure \cite{Mizukami}. In CeCu$_2$Si$_2$, we observe essentially no change in the low-temperature behavior of $\Delta\lambda(T/T_c)$ over a wide range of impurity scattering rates, indicating the robustness of the fully-gapped state against disorder. This provides strong evidence that the gap structure of CeCu$_2$Si$_2$ is non-sign changing $s$-wave state ($s_{++}$).

\begin{figure}[t!]
	\centering
	\includegraphics[clip,width=1.0\linewidth]{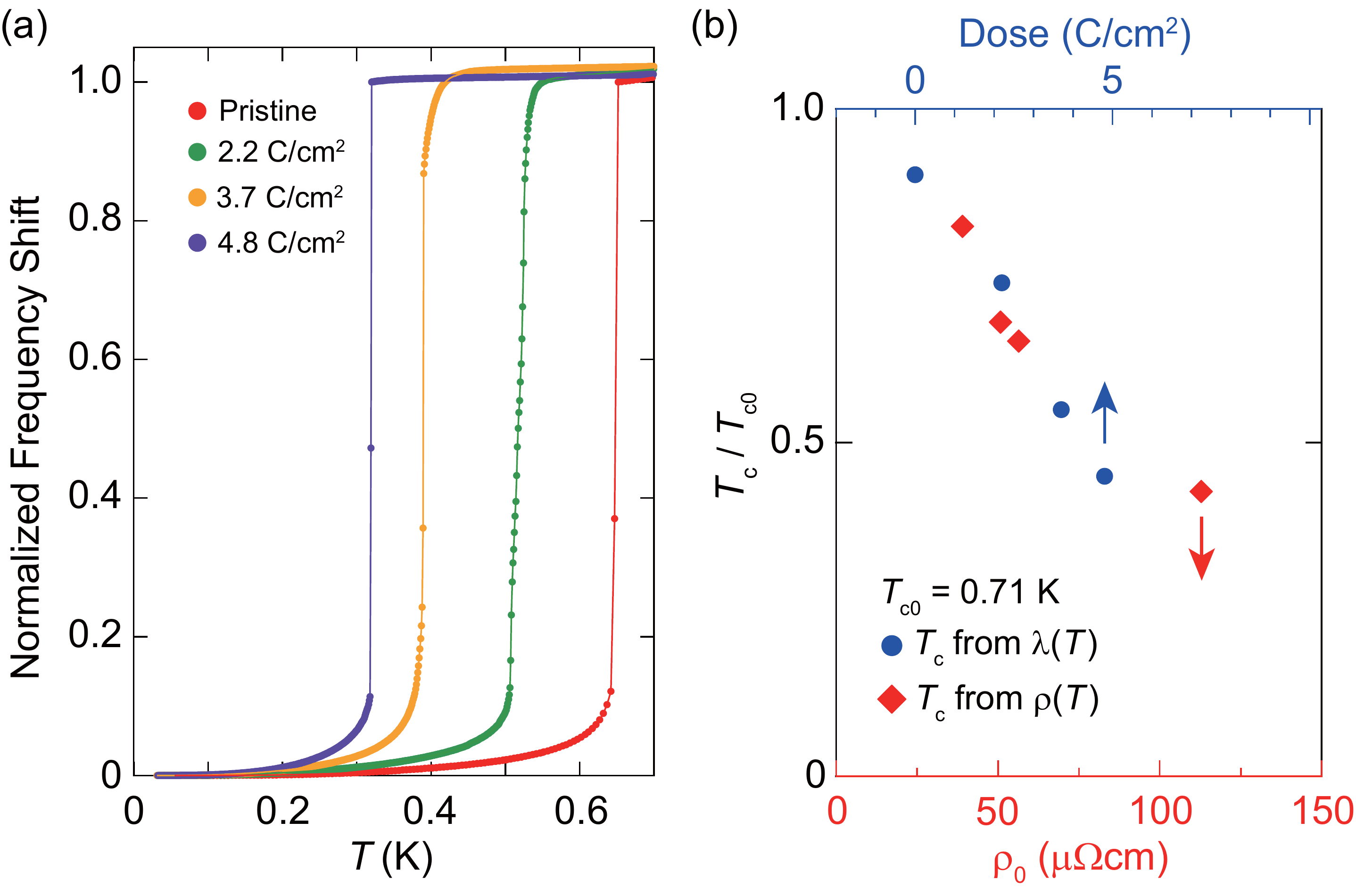}
	\caption{Superconducting transitions of CeCu$_2$Si$_2$ single crystals before and after electron irradiation. (a) Normalized frequency shift in the TDO measurements as a function of temperature below 0.7\,K for all measured samples. Samples were measured while warming to minimize ac field self-heating close to $T_c$.
	(b) Critical temperature $T_{{c}}$ defined as the midpoint of transition normalized by the clean-limit value $T_{{c0}}=0.71$\,K as a function of irradiation dose (upper axis). We also plot $T_c/T_{{c0}}$ as a function of residual resistivity $\rho_0$ taken from the reported resistivity data $\rho(T)$ \cite{Yamashita} (lower axis). The upper axis is adjusted to match the linear relation between the dose and $\rho_0$ \cite{Yamashita}.
	}
\end{figure}

High-quality single crystals of CeCu$_2$Si$_2$ were synthesized by the self flux method and characterized by x-ray diffraction \cite{Seiro}. The crystals were cut into samples with typical dimension about $350\times350\,\mu$m$^2$ (in the $ab$ plane) and  thickness about 50\,$\mu$m (along the $c$ axis). To introduce spatially homogeneous defects in a controllable way, we employed 2.5-MeV electron beam irradiation at the SIRIUS Pelletron linear accelerator operated by the Laboratorie des Solides Irradi{\'e}s (LSI) at {\'E}cole Polytechnique. This incident energy is sufficient to form vacancy-interstitial (Frenkel) pairs, which act as point-like defects. The attenuation distance of these irradiation electrons in CeCu$_2$Si$_2$  is about 2.7\,mm, which is much longer than our sample thickness. During the irradiation, the sample was kept at $\sim20$\,K by using a liquid hydrogen bath, which is important to prevent defect migration and clustering. The temperature dependent changes in the in-plane magnetic penetration depth $\Delta\lambda(T)$ were measured using the tunnel diode oscillator (TDO) technique operating at 14\,MHz \cite{carrington99} in a dilution refrigerator down to $\sim30$\,mK. The shift of the resonant frequency $\Delta f$ of the oscillator including the sample directly reflects the change in the magnetic penetration depth, $\Delta \lambda(T) = G\Delta f$. The geometric factor $G$ is determined from the geometry of the sample and the coil \cite{prozorov00}.

Figure\,1(a) shows the temperature dependence of the normalized frequency shifts for the pristine and irradiated samples with irradiation doses of 2.2, 3.7, and 4.8\,C/cm$^2$. The superconducting transition remains sharp after irradiation, indicating that the point-like defects are introduced uniformly. In Fig.\,1(b) we show the dose dependence of $T_c$ determined by the onset of the diamagnetic signal and compare this to the evolution of the $T_c$ and residual resistivity $\rho_0$ reported previously \cite{Yamashita}. Here the vertical axis is normalized by $T_{c0}=0.71$\,K, which is estimated from the linear extrapolation to the zero defect $\rho_0$ limit. These two independent results measured in different crystals are in good agreement, demonstrating that all irradiated samples are homogeneous.

\begin{figure}[t!]
	\centering
	\includegraphics[clip,width=0.9\linewidth]{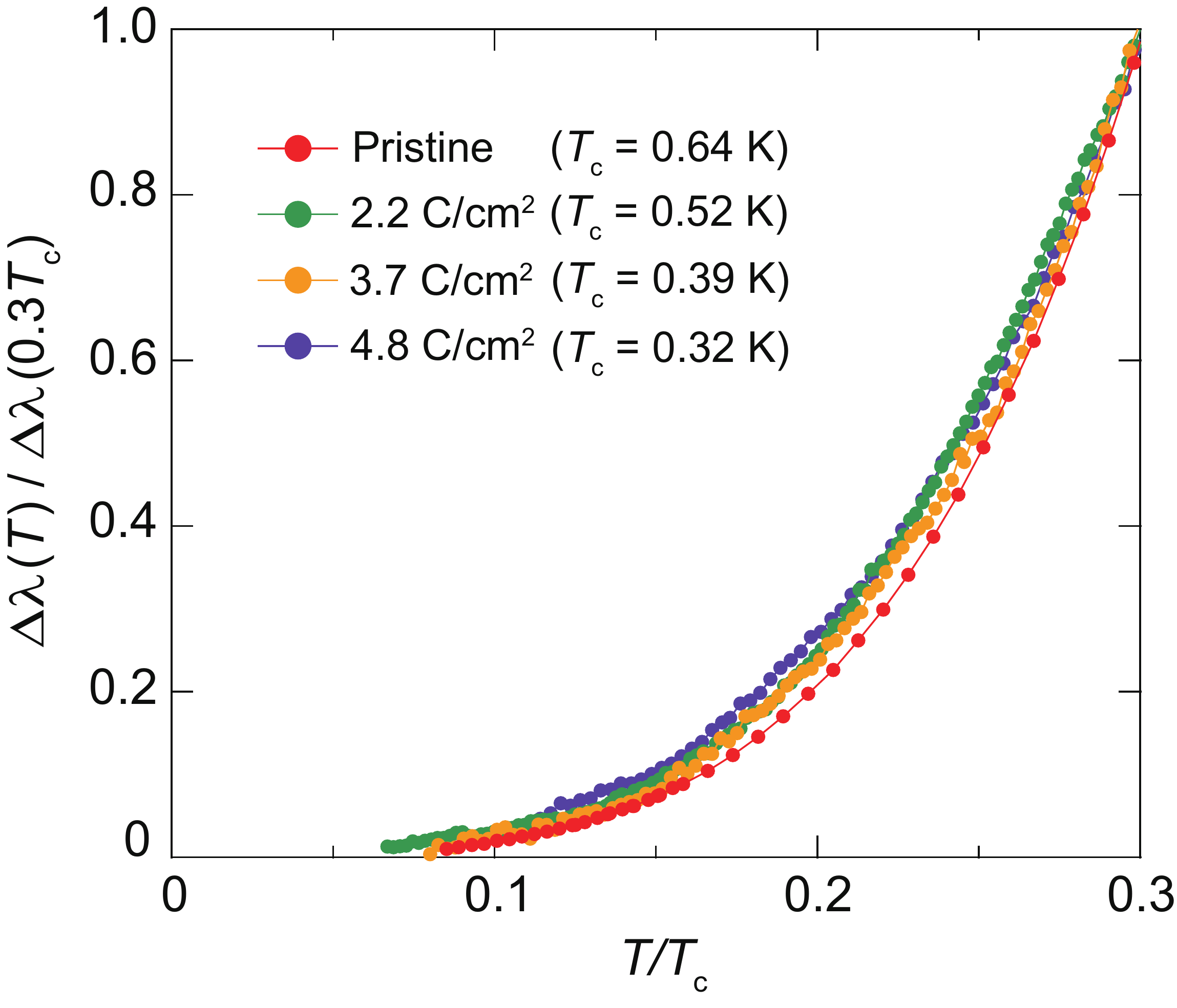}
	\caption{Temperature dependence of the change in the penetration depth $\Delta \lambda$ as a function of normalized temperature $T/T_{c}$. The origin of $\Delta \lambda$ at $T\to0$\,K is determined by the power-law fitting. The vertical axis is normalized by each value at $0.3 T_c$. Values of $\Delta \lambda(0.3T_c)$ are 29, 49, 38, and 30\,nm for doses of 0, 2.2, 3.7, and 4.8 C/cm$^2$, respectively. }
\end{figure}

Figure\,2 depicts the variations of penetration depth $\Delta\lambda$ as a function of $T/T_c$ for pristine and irradiated crystals, after normalization by their values of $\Delta\lambda(0.3T_c)$. We find no significant change in the temperature dependence of $\Delta\lambda$ at low temperatures, and all the curves almost collapse to a single one. This indicates that the introduced defects have essentially no effect on the low-energy quasiparticle excitations. The fact that we do not observe any evidence for a Curie-like upturn in $\Delta\lambda (T)$ down to the lowest temperature of $\sim30$\,mK even in most irradiated sample implies that introduced point-like defects are nonmagnetic in nature. Any magnetic impurities would result in a Curie-like upturn in the normal-state susceptibility, which would lead to an additional contribution $\Delta\lambda_m (T)$ to $\Delta\lambda (T)$, with
$\Delta \lambda_m \sim n\lambda (0)\mu_0\mu^2/3V_{\mathrm{cell}}k_B(T+\theta_N)$. $n$ and $\mu$ are respectively the density and the effective moment of the magnetic impurity \cite{Cooper,Malone,Mizukami}. We estimate about 2 vacancies per 1000 Ce atoms are formed per 1\,C/cm$^2$ electron irradiation, so the absence of the upturn in the 4.8\,C/cm$^2$ sample gives an upper limit of about $\mu\lesssim 0.5\,\mu_{\rm B}$ per defect. This is much smaller than the moment of $2.5 \mu_{\rm B}$ for a free Ce$^{3+}$ ion with total momentum $J=5/2$.

Having established the nonmagnetic nature of the defects, we analyze the low-temperature $\Delta\lambda(T)$ data to make a more quantitative analysis of the changes in the gap structure. We use two procedures: one is a fit to the power-law dependence $\Delta\lambda \propto(T/T_c)^n$ with a variable exponent $n$ and the other is a fit to the exponential dependence $\Delta\lambda \propto A T^{-1/2}\exp(-\Delta_{\rm min}/k_BT)$ with a variable minimum gap $\Delta_{\rm min}$. In both cases, we examine the changes in the fitting parameters as a function of the upper bound of fitted temperature range, $T_{\rm max}$. The obtained results for $n$ and $\Delta_{\rm min}$ are shown in Figs.\,3(a) and (b), respectively. For the power-law fitting procedure we find that all the data before and after irradiation give high exponent values $n>3$, which is far beyond the dirty-limit line-node case of $n=2$ exponent. This high-power dependence is practically indistinguishable from an exponential temperature dependence. The obtained gap values in the exponential fits are smaller than the BCS value of $1.76k_BT_c$, suggesting that the gap has strong $\bm{k}$ dependence with a large difference between minimum and maximum values.

\begin{figure}[t!]
	\centering
	\includegraphics[clip,width=1.0\linewidth]{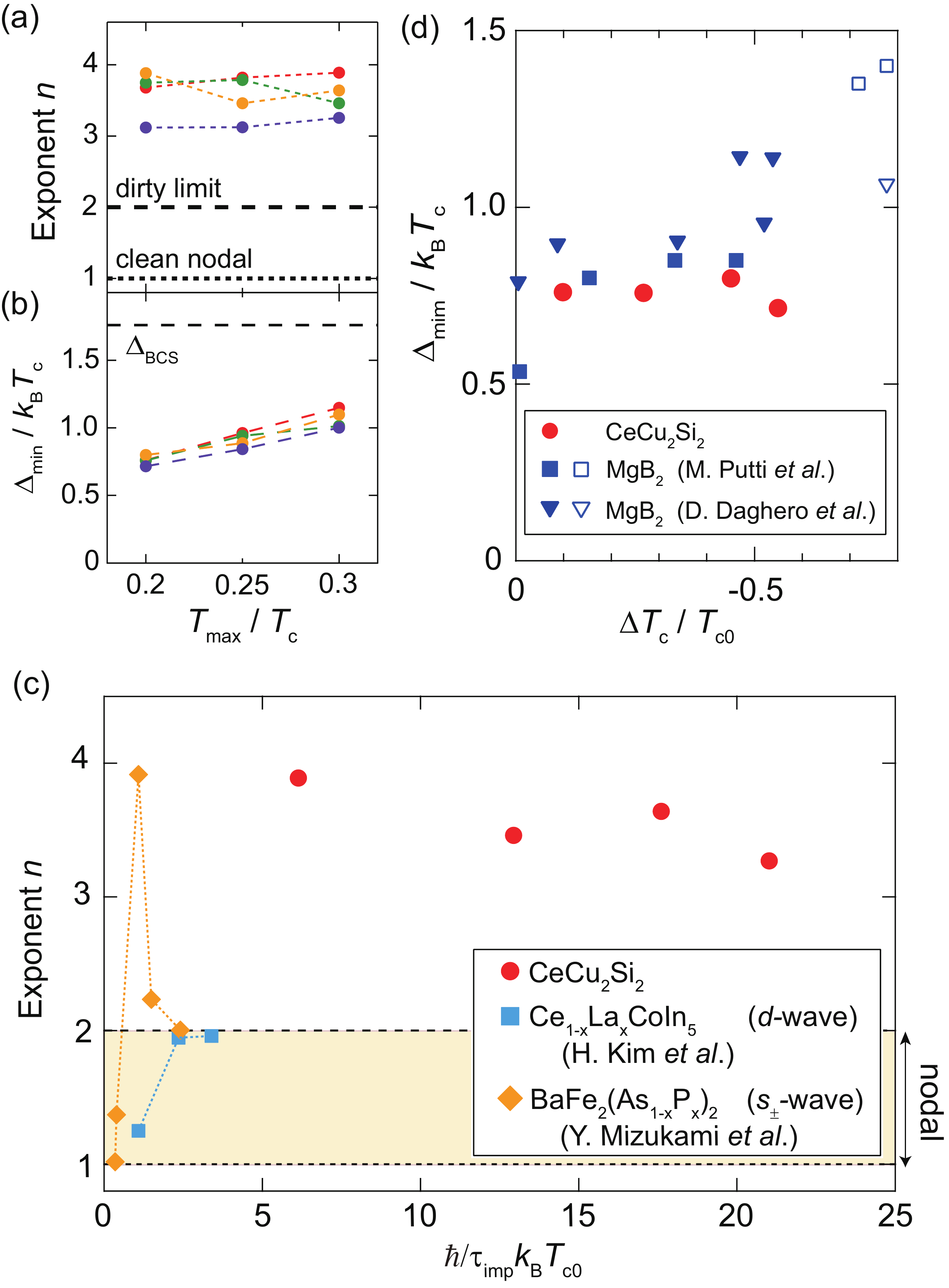}
	\caption{Disorder-induced changes of low-temperature penetration depth in  CeCu$_2$Si$_2$. (a) Exponent $n$ of a power-law fit of the experimental data up to $T_{\rm max}/T_c$. The colors for different doses are the same as in Fig.\,2. The dotted (dashed) line shows the clean (dirty) limit case of $n=1$ (2) in unconventional superconductors with line nodes. (b) Similar plot for minimum superconducting gap $\Delta_{\rm min}$ normalized by $k_BT_c$ obtained by the exponential fitting. (c) Exponent $n$ as a function of pair-breaking parameter $g = \hbar/\tau_{\mathrm{imp}}k_{\mathrm{B}}T_{c0}$, in comparison with those for  BaFe$_2$(As$_{1-x}$P$_x$)$_2$ \cite{Mizukami} and Ce$_{1-x}$La$_x$CoIn$_5$ \cite{Kim}. For Ce$_{1-x}$La$_x$CoIn$_5$, we use the values of $\lambda_{ab}(0)=200$\,nm and $\lambda_{c}(0)=280$\,nm \cite{Chia}, and $\rho_0$ is estimated from Ref.\,\cite{Petrovic}.
    (d) Normalized gap minima $\Delta_{\textrm{min}}/k_{B}T_c$ from the fit for $T_{\rm max}/T_c=0.2$ plotted against $\Delta T_c / T_{c0}$. For comparison, also plotted are the data for minimum gap in neutron-irradiated MgB$_2$ (closed squares and triangles), which change to a single gap (open symbols) for heavily irradiated samples \cite{Putti,Daghero}. }
\end{figure}

Our principal finding is the robustness of the fully-gapped superconductivity against disorder in CeCu$_2$Si$_2$. This is most clearly demonstrated by plotting the exponent $n$ in the power-law fit as a function of pair-breaking parameter $g = \hbar/\tau_{\mathrm{imp}}k_{\mathrm{B}}T_{c0}$ in Fig.\,3(c), in which we compare with the typical results for $d$-wave Ce$_{1-x}$La$_x$CoIn$_5$ \cite{Kim} and for $s_{\pm}$-wave BaFe$_2$(As$_{1-x}$P$_x$)$_2$ \cite{Mizukami}. The impurity scattering time $\tau_{\mathrm{imp}}$ is calculated with $\tau_{\mathrm{imp}}=\mu_0 \lambda_{ab}\lambda_c / \rho_0$. In La-substituted CeCoIn$_5$ the exponent increases with impurity scattering and saturates at $n\approx2$, which is consistent with the gapless state expected theoretically in the dirty $d$-wave superconductors \cite{Hirschfeld} and established experimentally for Zn-substituted YBa$_2$Cu$_3$O$_7$ \cite{Bonn}. In optimal BaFe$_2$(As$_{1-x}$P$_x$)$_2$, which has a $T$-linear behavior due to the accidental line nodes in the clean limit \cite{Hashimoto}, the exponent initially shows a large increase from $n\approx 1$ to $n\approx 4$, indicating the lifting of nodes by the impurity-induced averaging effect of the $\bm{k}$ dependence, which occurs only when the nodes are not symmetry protected. Further irradiation yields a decrease of $n$ toward the gapless value of 2, demonstrating the creation of the low-energy states that are expected only for sign-changing cases. These results established a nodal $s_{\pm}$-wave state in this iron pnictide. Thus in both $d$-wave and $s_{\pm}$-wave cases, a gapless state with the exponent $n=2$ appears for pair-breaking parameter $g$ of the order of unity. In stark contrast, our data for CeCu$_2$Si$_2$ reveal that the exponent remains high ($n>3$) even when $g$ exceeds 20, which clearly indicates the absence of impurity-induced low-energy states, evidencing no sign change in $\Delta(\bm{k})$.

The minimum superconducting gap size $\Delta_{\rm min}$ normalized by $k_BT_c$ shows no appreciable change against the relative suppression of the transition temperature $\Delta T_c/T_{c0}$ as shown in Fig.\,3(d). At first glance this appears counterintuitive because the gap averaging effect due to impurity scattering might be expected to lead to an increase of the minimum $\Delta/T_c$. However similar behavior is also observed in the protypical two gap superconductor MgB$_2$, where following an initial increase, $\Delta_{\rm min}/T_c$  remains unchanged in a wide $T_c$ suppression range up to about half of $T_{c0}$ \cite{Putti,Daghero}.

The anisotropic nature of the gap structure in CeCu$_2$Si$_2$ can be seen in the full temperature dependence of the normalized superfluid density $\rho_s (T) = \lambda^2(0)/\lambda^2(T)$. To calculate $\rho_s (T)$, we need the value of $\lambda(0)$ for each sample, which cannot be determined directly by using the TDO technique.  So instead we have estimated $\lambda(0)$ from the lower critical field $H_{c1}(T)$ measured by micro Hall-probe magnetometry as described in Ref.\,\cite{SM}. To minimize errors due to geometrical demagnetization factors we measured the same sample both before and after irradiation with a dose of 1.9\,C/cm$^2$, which reduced $T_c$ from 0.64\,K to 0.52\,K.  For the irradiated sample we found $\mu_0 H_{c1}=0.9\pm 0.1$\,mT at 100\,mK compared to $\mu_0 H_{c1}(0)=1.8\pm 0.1$\,mT in the pristine sample \cite{Yamashita}. From this we estimate that $\lambda(0)$ is increased from $700 \pm 50$\,nm for the unirradiated sample to $1100 \pm 100$\,nm for the irradiated one. An increase in $\lambda(0)$ upon irradiation is expected because the effective penetration depth depends on the mean free path $\ell$ of quasiparticles. For the unirradiated sample we estimated the in-plane mean free paths and coherence lengths to be $\ell = 3.0$\,nm and $\xi_{ab}=4.7$\,nm respectively so the sample is between the clean and dirty limits. Then from the change in $\rho_0$ we would expect $\ell$ to decrease by a factor 2 for this irradiation level, pushing the sample closer to the dirty limit and thus increasing $\lambda(0)$. In the dirty limit, $H_{c1}\propto\lambda^{-2}$ is expected to be proportional to $\ell\propto1/\rho_0$, which appears to hold as shown in the inset of Fig.\,4. From this relation we estimate $\lambda(0)\approx 1390$\,nm for the most irradiated sample (4.8\,C/cm$^2$). Figure 4 displays the extracted $\rho_s(T)$ curves before and after irradiation, which again show the robustness of flat temperature dependence at low temperatures indicating the absence of the low-energy states. The multi-gap or strong $\bm{k}$-dependent nature of $\Delta(\bm{k})$ manifests itself in the concave curvature near $T_c$. Contrary to the case of MgB$_2$ \cite{Fletcher}, however, a simple two-gap model does not fit the $\rho_s(T)$ data very well. Possible reasons for this include significant interband scattering and largely varying $\Delta(\bm{k})$ for each band. For the irradiated sample, the concave curvature of $\rho_s(T)$ near $T_c$ is less pronounced and the curve becomes closer to the single-gap $s$-wave one, which is consistent with the reduced anisotropy of $\Delta(\bm{k})$ by impurity scattering. 

\begin{figure}[t!b]
	\centering
	\includegraphics[clip,width=\linewidth]{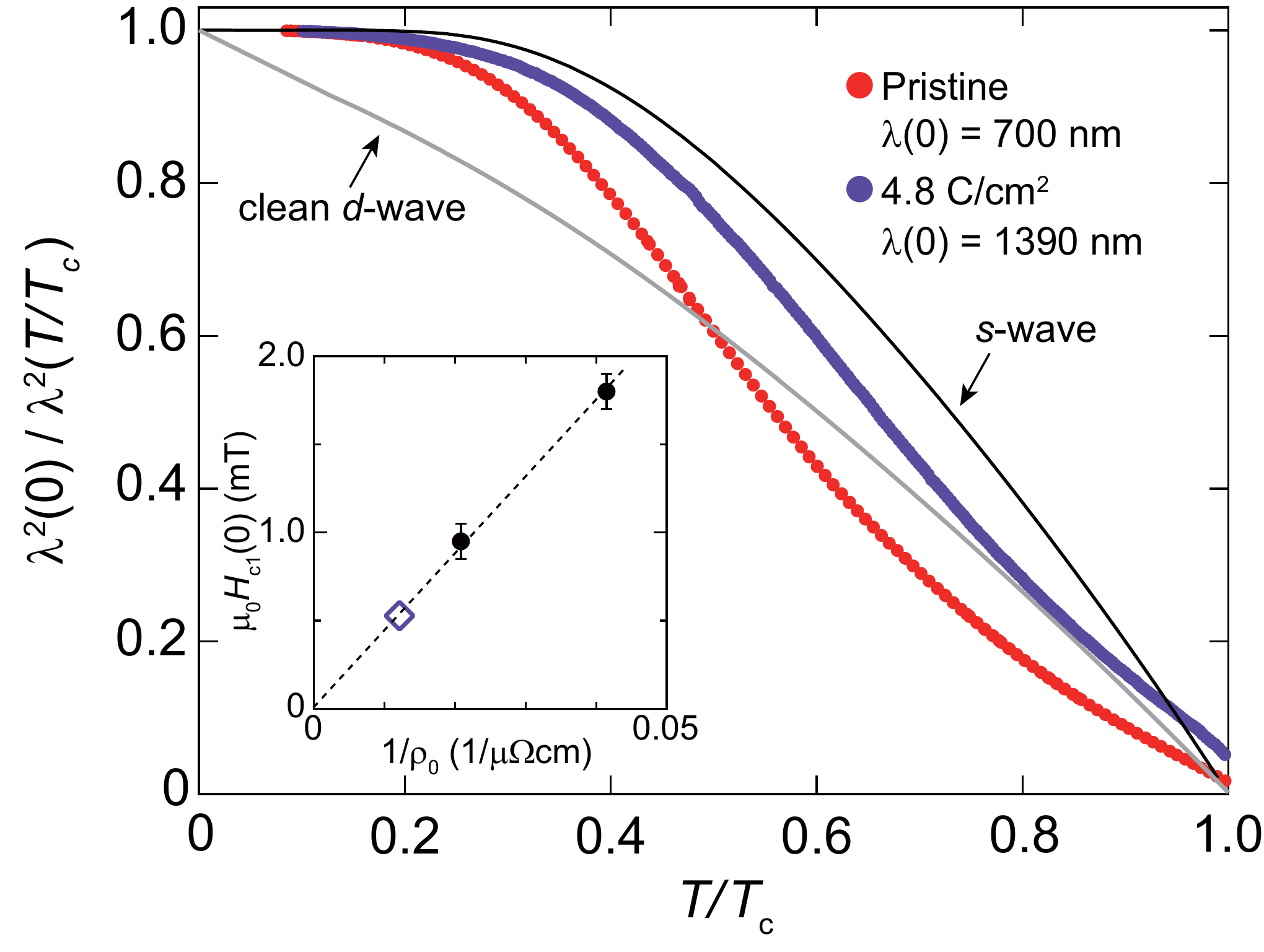}
	\caption{Temperature dependence of normalized superfluid density $\rho_s (T)=\lambda^2(0)/\lambda^2(T)$ in CeCu$_2$Si$_2$ before and after irradiation. 	Black and gray line are the theoretical temperature dependence of $\rho_s(T)$ in the conventional $s$-wave (BCS) and clean-limit $d$-wave cases, respectively.
	Inset shows the measured lower critical field $H_{c1}$ (closed circles) plotted against the inverse of residual resistivity $1/\rho_0$. The dashed line evidences a linear relation, from which $H_c1(0)$ value for 4.8\,C/cm$^2$ (open diamond) and thus the corresponding $\lambda(0)$ value are estimated.
}
\end{figure}

In summary, systematic measurements of magnetic penetration depth $\lambda(T)$ in electron-irradiated single crystals of CeCu$_2$Si$_2$ show that nonmagnetic impurity scattering does not induce any low-energy quasiparticle excitations. This provides bulk evidence for the absence of a sign change in the gap function in the superconducting state of this heavy-fermion superconductor. 
The $s_{++}$-wave state inferred in this study is generally a manifestation of on-site attractive interactions, but how this can overcome the strong Coulomb repulsion in such a strongly correlated electron system calls for new theoretical approaches beyond the wide-spread spin-fluctuation based unconventional mechanism of superconductivity.
Very recent calculations show that in the vicinity of magnetic quantum critical point, the orbital fluctuations may lead to $s_{++}$-wave superconductivity \cite{Tazai}. Indeed, the importance of orbital degrees of freedom has been pointed out in several aspects for some Ce-based materials including Ce$M$$_2$Si$_2$, where $M$ is a transition metal element \cite{Hattori,Ren,Dong} . 
Thus the relationship between the orbital effects and gap symmetry in heavy-fermion superconductors deserves further studies. 

We thank H. Ikeda and H. Kontani for fruitful discussion. We also thank the SIRIUS team, O. Cavani, B. Boizot, V. Metayer, and J. Losco, for running electron irradiation at Laboratorie des Solides Irradi{\'e}s (LSI) in {\'E}cole Polytechnique [supported by EMIR network, proposal Nos. 16-0398, 16-9513, and 17-1353]. 
This work was supported by Grants-in-Aid for Scientific Research(KAKENHI) (No. 25220710, No. 15H02106) and Grants-in-Aid for Scientific Research on Innovative Areas ``Topological Materials Science" (No. 15H05852) from Japan Society for the Promotion of Science (JSPS), and by the UK Engineering and Physical Sciences Research Council (grant no.EP/L025736/1 and EP/L015544/1).

\renewcommand{\figurename}{Figure S$\!\!$}
\renewcommand{\tablename}{Table S$\!\!$}
\renewcommand{\theequation}{S\arabic{equation}}
\setcounter{figure}{0}

\section{Supplemental Material}

\subsection{Lower Critical Field}

\begin{figure} [b]
	\includegraphics[width=\linewidth]{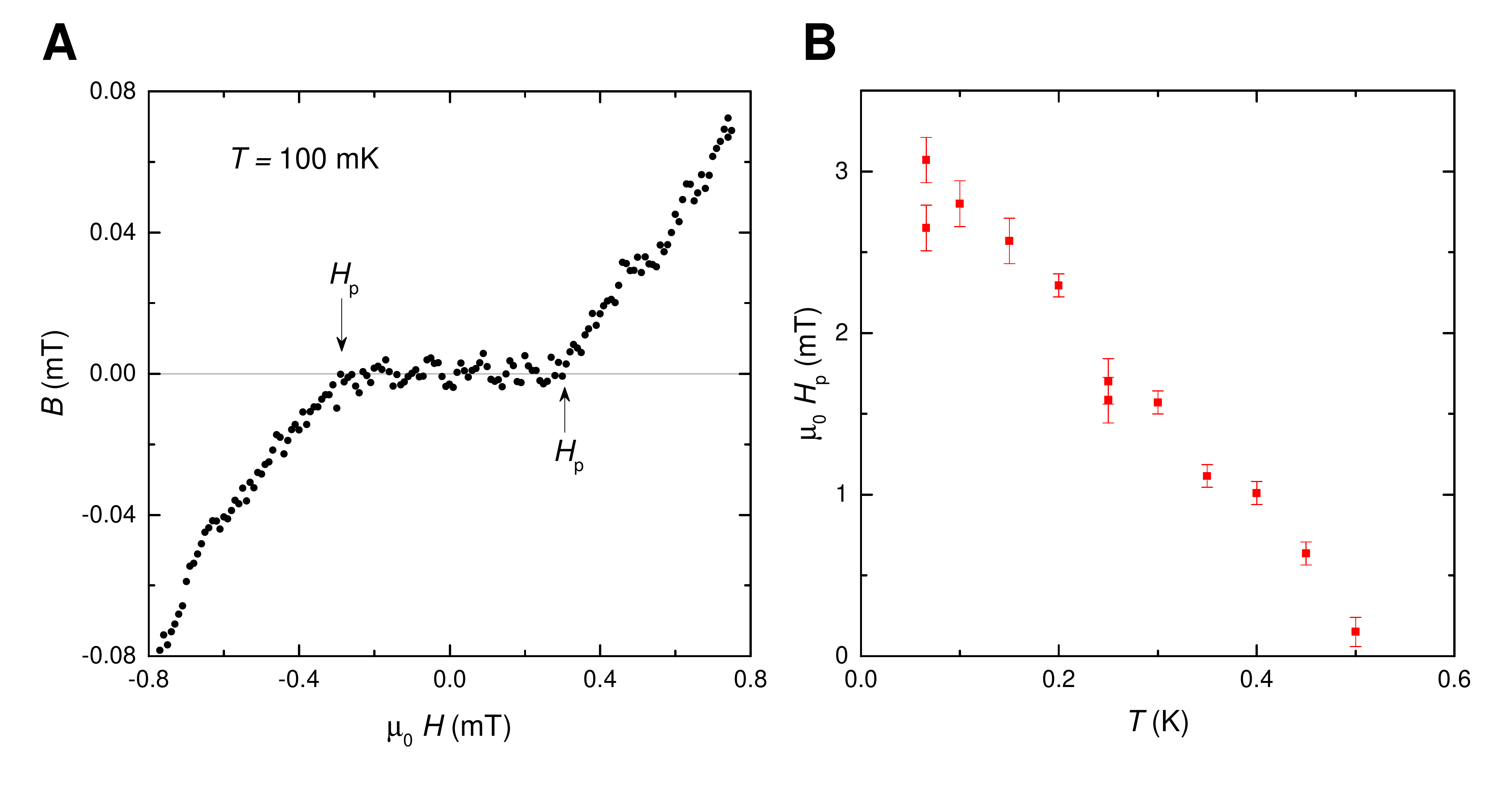}
	\vspace{-4mm}
	\caption{{Lower critical field measurements of irradiated CeCu$_2$Si$_2$.} (a) Magnetic induction ($B$) versus applied field ($H$) as measured by a Hall sensor located below the middle of the sample that has an irradiation dose of 1.9 C cm$^{-2}$. The figure shows two sweeps - one for increasing field and another for decreasing field. The sample was cooled through $T_{\mathrm{c}}$ in zero field before each sweep. A linear background has been subtracted to account for the incomplete shielding of the sample and the values for $H_{\mathrm{p}}$ are indicated. (b) The field of first flux penetration $H_{\mathrm{p}}$ versus temperature. The error bars reflect the uncertainty in determining $H_{\mathrm{p}}$.}
	\label{fig:S1}
\end{figure}

In order to estimate the absolute value of the penetration depth in the pristine and irradiated samples, we employ a technique using a micro-Hall array to determine the lower critical field $H_{\mathrm{c1}}$ of the samples.
The sample is placed on top of an array of Hall sensors, where each Hall sensor measures the perpendicular component of the magnetic induction $B$ through the active area as a function of an applied field $H$. The sample is cooled in zero-field, with the vacuum can of the refrigerator shielded in mu-metal to reduce the effect of the earth's field.  A small coil inside the can provides the magnetic field.

At a given temperature, the applied field, oriented perpendicular to the Hall sensors and parallel to the $c$-axis of the sample, is increased from zero to the maximum value (positive or negative).  At low field, there is an increase in $B$ with $H$ due to the incomplete shielding of the Hall sensors by the sample. At a well defined field, the magnetic induction increases sharply, indicating the field $H_{\mathrm{p}}$ at which flux has entered the sample. An example sweep is shown in Fig.\,S\ref{fig:S1}(a). A linear background is subtracted from the data to account for the incomplete shielding below $H_{\mathrm{p}}$, and the values of $H_{\mathrm{p}}$ for each sweep direction are indicated. After each sweep, the sample is warmed above $T_\mathrm{c}$ and cooled again in zero-field. The overall temperature dependence is given in Fig.\,S\ref{fig:S1}(b).

The field at which flux enters the sample is not equal to the lower critical field $H_{\mathrm{c1}}$, but it is the lower critical field reduced by a factor related to the demagnetising effects due to the geometry of the sample. The lower critical field is related to the penetration field by the relation given by Brandt for a strip \cite{Brandt2}:
\begin{equation}
H_{\mathrm{c1}} = \frac{H_\mathrm{p}}{\tanh\sqrt{0.36 \, c / a}} \, ,
\end{equation}
where we take $a$ as the shorter of the two in-plane dimensions and $c$ is the thickness in the $c$-axis direction. Although this expression was calculated for a strip of infinite third dimension, it has been found to provide a good description of samples even if the two in-plane dimensions are approximately equal (see for example Ref.\ \cite{Putzke2} where consistent results were found when the aspect ratio of a sample was varied by cutting). For the sample with an irradiation dose of 1.9 C cm$^{-2}$ (dimensions $0.29 \times 0.40 \times 0.09$ mm$^3$) we find a value of $\mu_0 H_{\mathrm{c1}} = 0.9 \pm 0.1$ mT at $T = 100$\,mK. The penetration depth is then determined from the Ginzburg-Landau equation
\begin{equation}
H_\mathrm{c1} = \frac{\phi_0}{4 \pi \lambda^2}\left \lbrack \ln\left(\frac{\lambda}{\xi}\right)+0.5\right\rbrack \, ,
\end{equation}
where $\phi_0$ is the flux quantum and $\xi$ is the coherence length. 
Using $\xi = 4.7$\,nm found for the pristine sample \cite{Yamashita2}, gives a value of $\lambda = 1100 \pm 100$\,nm at $T=100$\,mK which within the uncertainty we take as being equal to the zero temperature value.


\begin{thebibliography}{99}
	\providecommand \doibase [0]{http://dx.doi.org/}%
	
	\bibitem{Yamashita}
	{T.}~{Yamashita}, {T.}~{Takenaka}, {Y.}~{Tokiwa}, {J.~A.}~{Wilcox}, {Y.}~{Mizukami}, {D.}~{Terazawa}, {Y.}~{Kasahara}, {S.}~{Kittaka}, {T.}~{Sakakibara}, {M.}~{Ko{\'n}czykowski}, {S.}~{Seiro}, {H.~S.}~{Jeevan}, {C.}~{Geibel}, {C.}~{Putzke}, {T.}~{Onishi}, {H.}~{Ikeda}, {A.}~{Carrington}, {T.}~{Shibauchi},\ and\ {Y.}~{Matsuda}, Sci. Adv. \textbf{3}, e1601667 (2017).
	
	\bibitem{Steglich}
	{F.}~{Steglich}, {J.}~{Aarts}, {C.~D.}~{Bredl}, {W.}~{Lieke}, {D.}~{Meschede}, {W.}~{Franz},\ and\ {H.}~{Sch\"afer},\
	\href {\doibase 10.1103/PhysRevLett.43.1892}
	{\bibfield  {journal}
		{\bibinfo  {journal}{Phys. Rev. Lett.}}
		\textbf {\bibinfo {volume} {43}},\
		\bibinfo {pages}{1892}
		(\bibinfo {year} {1979})}.
	
	\bibitem{Ishida}
	{K.}~{Ishida}, {Y.}~{Kawasaki}, {K.}~{Tabuchi}, {K.}~{Kashima},  {Y.}~{Kitaoka}, {K.}~{Asayama},{C.}~{Geibel},\ and\ {F.}~{Steglich},\
	\href {\doibase 10.1103/PhysRevLett.82.5353}
	{\bibfield  {journal}
		{\bibinfo  {journal} {Phys. Rev. Lett.}}
		\textbf{\bibinfo {volume} {82}},\
		\bibinfo {pages} {5353}
		(\bibinfo {year}{1999})}.
	
	\bibitem{Fujiwara}
	{K.}~{Fujiwara}, {Y.}~{Hata}, {K.}~{Kobayashi}, {K.}~{Miyoshi}, {J.}~{Takeuchi}, {Y.}~{Shimaoka}, {H.}~{Kotegawa},  {T.~C.}~{Kobayashi}, {C.}~{Geibel},\ and\ {F.}~{Steglich},\
	\href {\doibase  10.1143/JPSJ.77.123711}
	{\bibfield  {journal}
		{\bibinfo  {journal} {J. Phys. Soc. Jpn.}}
		\textbf {\bibinfo {volume} {77}},\
		\bibinfo {pages} {123711}
		(\bibinfo {year} {2008})}.

	\bibitem{Arndt}
{J.}~{Arndt}, {O.}~{Stockert}, {K.}~{Schmalzl}, {E.}~{Faulhaber}, {H.~S.}~{Jeevan}, {C.}~{Geibel}, {W.}~{Schmidt}, {M.}~{Loewenhaupt},\ and\ {F.}~{Steglich},\
\href {\doibase  10.1103/PhysRevLett.106.246401}
{\bibfield  {journal} {\bibinfo  {journal} {Phys. Rev. Lett.}}
	\textbf {\bibinfo {volume} {106}},\
	\bibinfo {pages}{246401}
	(\bibinfo {year} {2011})}.
	
	\bibitem{Stockert}
	{O.}~{Stockert}, {J.}~{Arndt}, {E.}~{Faulhaber}, {C.}~{Geibel}, {H.}~{Jeevan}, {S.}~{Kirchner}, {M.}~{Loewenhaupt}, {K.}~{Schmalzl}, {W.}~{Schmidt}, {Q.}~{Si},\ and\ {F.}~{Steglich},\
	{\bibfield  {journal} {\bibinfo  {journal} {Nat. Phys.}}
		\textbf {\bibinfo {volume} {7}},\
		\bibinfo {pages} {119}
		(\bibinfo {year} {2011})}.
	
	\bibitem{Eremin}
	{I.}~{Eremin}, {G.}~{Zwicknagl}, {P.}~{Thalmeier},\ and\ {P.}~{Fulde},\
	\href {\doibase 10.1103/PhysRevLett.101.187001}
	{\bibfield  {journal} {\bibinfo {journal} {Phys. Rev. Lett.}}
		\textbf {\bibinfo {volume} {101}},\
		\bibinfo{pages} {187001}
		(\bibinfo {year} {2008})}.
	
	\bibitem{Kittaka}
	{S.}~{Kittaka}, {Y.}~{Aoki}, {Y.}~{Shimura}, {T.}~{Sakakibara}, {S.}~{Seiro}, {C.}~{Geibel}, {F.}~{Steglich}, {H.}~{Ikeda},\ and\ {K.}~{Machida},\
	\href {\doibase 10.1103/PhysRevLett.112.067002}
	{\bibfield  {journal} {\bibinfo  {journal} {Phys. Rev. Lett.}}
		\textbf {\bibinfo {volume} {112}},\
		\bibinfo {pages} {067002}
		(\bibinfo {year} {2014})}.
	
	\bibitem{Ikeda}
	{H.}~{Ikeda}, {M.-T.}~{Suzuki},\ and\ {R.}~{Arita},\
	\href {\doibase 10.1103/PhysRevLett.114.147003}
	{\bibfield  {journal} {\bibinfo {journal} {Phys. Rev. Lett.}}
		\textbf {\bibinfo {volume} {114}},\
		\bibinfo {pages} {147003}
		(\bibinfo {year} {2015})}.
	
		\bibitem{Adrian}
	{G.}~{Adrian}\ and\ {H.}~{Adrian},\
	\href {http://stacks.iop.org/0295-5075/3/i=7/a=008}
	{\bibfield{journal} {\bibinfo  {journal} {Europhys. Lett.}}
		\textbf{\bibinfo {volume} {3}},\
		\bibinfo {pages} {819}
		(\bibinfo {year}{1987})}.
	
		\bibitem{Poppe}
	{U.}~{Poppe},\
	\href {\doibase http://dx.doi.org/10.1016/0304-8853(85)90243-4}
	{\bibfield  {journal} {\bibinfo  {journal} {J. Magn. Magn. Mater.}}
		\textbf {\bibinfo {volume} {52}},\
		\bibinfo {pages} {157}
		(\bibinfo {year} {1985})}.
	
	\bibitem{Sumiyama}
	{A.}~{Sumiyama}, {N.}~{Miyakawa}, {Y.}~{Ushida}, {G.}~{Motoyama}, {A.}~{Yamaguchi},\ and\ {Y.}~{Oda},\
	\href  {http://stacks.iop.org/1742-6596/273/i=1/a=012086}
	{\bibfield  {journal}  {\bibinfo  {journal} {J. Phys: Conf. Ser.}}
		\textbf{\bibinfo {volume} {273}},\
		\bibinfo {pages} {012086}
		(\bibinfo {year}{2011})}.
		
	\bibitem{Wang}
	{Y.}~{Wang}, {A.}~{Kreisel}, {P.~J.}~{Hirschfeld},\ and\ {V.}~{Mishra},\
	\href {\doibase 10.1103/PhysRevB.87.094504}
	{\bibfield  {journal} {\bibinfo {journal} {Phys. Rev. B}}
		\textbf {\bibinfo {volume} {87}},\
		\bibinfo {pages} {094504}
		(\bibinfo {year} {2013})}.
	

	\bibitem{Mizukami}%
	{Y.}~{Mizukami}, {M.}~{Ko{\'n}czykowski}, {Y.}~{Kawamoto}, {S.}~{Kurata}, {S.}~{Kasahara}, {K.}~{Hashimoto}, {V.}~{Mishra}, {A.}~{Kreisel}, {Y.}~{Wang}, {P.~J.}~{Hirschfeld}, {Y.}~{Matsuda},\ and\ {T.}~{Shibauchi},\
	{\bibfield  {journal} {\bibinfo  {journal} {Nat. Commun.}}
		\textbf {\bibinfo {volume} {5}},
		\bibinfo {pages} {5657}
		(\bibinfo {year} {2014})}.
	
	\bibitem{Seiro}
	{S.}~{Seiro}, {M.}~{Deppe}, {H.}~{Jeevan}, {U.}~{Burkhardt},\ and\ {C.}~{Geibel},\
	{\bibfield  {journal} {\bibinfo  {journal} {Phys. Status Solidi B}}
		\textbf {\bibinfo {volume} {247}},\
		\bibinfo {pages} {614}
		(\bibinfo {year} {2010})}.
	
	  \bibitem{carrington99}
	A. Carrington, R. W. Giannetta, J. T. Kim, and J. Giapintzakis
	{\bibfield{journal} {\bibinfo  {journal} {Phys. Rev. B}}
		\textbf{\bibinfo {volume} {59}},\
		\bibinfo {pages} {R14173}
		(\bibinfo {year}{1999})}.
	
	\bibitem{prozorov00}
	R. Prozorov, R. W. Giannetta, A. Carrington, and F. M. Araujo-Moreira
	{\bibfield{journal} {\bibinfo  {journal} {Phys. Rev. B}}
		\textbf{\bibinfo {volume} {62}},\
		\bibinfo {pages} {115}
		(\bibinfo {year}{2000})}.
	
	\bibitem{Cooper}
	{J.~R.}~{Cooper},\
	\href {\doibase 10.1103/PhysRevB.54.R3753}
	{\bibfield  {journal} {\bibinfo  {journal} {Phys. Rev. B}}
		\textbf {\bibinfo {volume} {54}},\
		\bibinfo {pages} {R3753}
		(\bibinfo {year} {1996})}.
	%
	\bibitem{Malone}
	{L.}~{Malone}, {J.~D.}~{Fletcher}, {A.}~{Serafin}, {A.}~{Carrington}, {N.~D.}~{Zhigadlo}, {Z.}~{Bukowski}, {S.}~{Katrych},\ and\ {J.}~{Karpinski},\
	\href {\doibase 10.1103/PhysRevB.79.140501}
	{\bibfield  {journal} {\bibinfo  {journal} {Phys. Rev. B}}
		\textbf {\bibinfo {volume} {79}},\
		\bibinfo {pages} {140501}
		(\bibinfo {year} {2009})}.
	
	\bibitem{Kim}
	{H.}~{Kim}, {M.~A.}~{Tanatar}, {R.}~{Flint}, {C.}~{Petrovic}, {R.}~{Hu}, {B.~D.}~{White}, {I.~K.}~{Lum}, {M.~B.}~{Maple},\ and\ {R.}~{Prozorov},\
	\href {\doibase 10.1103/PhysRevLett.114.027003}
	{\bibfield {journal} {\bibinfo  {journal} {Phys. Rev. Lett.}}
		\textbf {\bibinfo {volume} {114}},\
		\bibinfo {pages} {027003}
		(\bibinfo {year} {2015})}.
	
	
	\bibitem{Chia}
	{E.~E.~M.}~{Chia}, {D.~J.}~{Van~Harlingen}, {M.~B.}~{Salamon}, {B.~D.}~{Yanoff}, {I.}~{Bonalde},\ and\ {J.~L.}~{Sarrao},\
	\href {\doibase 10.1103/PhysRevB.67.014527}
	{\bibfield  {journal} {\bibinfo {journal} {Phys. Rev. B}}
		\textbf {\bibinfo {volume} {67}},\
		\bibinfo {pages} {014527}
		(\bibinfo {year} {2003})}.
	
	\bibitem{Petrovic}
	{C.}~{Petrovic}, {S.~L.}~{Bud'ko}, {V.~G.}~{Kogan},\ and\ {P.~C.}~{Canfield},\
	\href {\doibase 10.1103/PhysRevB.66.054534}
	{\bibfield  {journal} {\bibinfo {journal} {Phys. Rev. B}}
		\textbf {\bibinfo {volume} {66}},\
		\bibinfo {pages} {054534}
		(\bibinfo {year} {2002})}.
	
	\bibitem{Putti}
{M.}~{Putti}, {M.}~{Affronte}, {C.}~{Ferdeghini}, {P.}~{Manfrinetti}, {C.}~{Tarantini},\ and\ {E.}~{Lehmann},\
\href {\doibase 10.1103/PhysRevLett.96.077003}
{\bibfield  {journal} {\bibinfo  {journal} {Phys. Rev. Lett.}}
	\textbf {\bibinfo {volume} {96}},\
	\bibinfo {pages} {077003}
	(\bibinfo {year} {2006})}.




\bibitem{Daghero}
{D.}~{Daghero}, {A.}~{Calzolari}, {G.~A.}~{Ummarino}, {M.}~{Tortello}, {R.~S.}~{Gonnelli}, {V.~A.}~{Stepanov}, {C.}~{Tarantini}, {P.}~{Manfrinetti},\ and\ {E.}~{Lehmann},\
\href {\doibase 10.1103/PhysRevB.74.174519}
{\bibfield  {journal} {\bibinfo  {journal} {Phys. Rev. B}}
	\textbf {\bibinfo {volume} {74}},\
	\bibinfo {pages} {174519}
	(\bibinfo {year} {2006})}.

	\bibitem{Hirschfeld}
	{P.~J.}~{Hirschfeld}\ and\ {N.}~{Goldenfeld},\
	\href {\doibase 10.1103/PhysRevB.48.4219}
	{\bibfield  {journal} {\bibinfo  {journal} {Phys. Rev. B}}
		\textbf {\bibinfo {volume}  {48}},\
		\bibinfo {pages} {4219}
		(\bibinfo {year} {1993})}.
	
	\bibitem{Bonn}
{D.~A.}~{Bonn}, {S.}~{Kamal}, {K.}~{Zhang}, {R.}~{Liang}, {D.~J.}~{Baar}, {E.}~{Klein},\ and\ {W.~N.}~{Hardy},\
\href {\doibase  10.1103/PhysRevB.50.4051}
{\bibfield  {journal} {\bibinfo  {journal} {Phys. Rev. B}}
	\textbf {\bibinfo {volume} {50}},\
	\bibinfo {pages} {4051}
	(\bibinfo {year} {1994})}.

  \bibitem{Hashimoto}
  K. Hashimoto, M. Yamashita, S. Kasahara, Y. Senshu, N. Nakata, S. Tonegawa, K. Ikada, A. Serafin, A. Carrington, T. Terashima, H. Ikeda, T. Shibauchi, and Y. Matsuda,
  Phys. Rev. B {\bf 81}, 220501(R) (2010).

\bibitem{SM} See Supplemental Material at http://link.aps.org/supplemental/10.1103/PhysRevLett.*** for details of the $H_{c1}$ measurements, which includes Refs.\,\cite{Brandt,Putzke}.

\bibitem{Brandt} E. H. Brandt, {Phys. Rev. B}
\textbf{60}, 11939 (1999).

\bibitem{Putzke} C. Putzke, P. Walmsley, J.D. Fletcher, L. Malone, D. Vignolles, C. Proust, S. Badoux, P. See, H. E. Beere,
D. A. Ritchie, S. Kasahara, Y. Mizukami, T. Shibauchi, Y. Matsuda, and A. Carrington, {Nat. Commun.} \textbf{5}, 5679 (2014).

\bibitem{Fletcher} J. D. Fletcher, A. Carrington, O. J. Taylor, S. M. Kazakov, and J. Karpinski, Phys. Rev. Lett. {\bf 95}, 097005 (2005).

\bibitem{Tazai} R. Tazai, Y. Yamakawa, M. Tsuchiizu, and H. Kontani, J. Phys. Soc. Jpn. {\bf 86}, 073703 (2017).  

\bibitem{Hattori} K. Hattori, J. Phys. Soc. Jpn. {\bf 79}, 114717 (2010). 

\bibitem{Ren} Z. Ren, L. V. Pourovskii, G. Giriat, G. Lapertot, A. Georges, and D. Jaccard, Phys. Rev. X {\bf 4}, 031055 (2014).

\bibitem{Dong} R. Dong, X. Wan, X. Dai, and S. Y. Savrasov, Phys. Rev. B {\bf 89}, 165122 (2014).

 \end{thebibliography}

\begin{thebibliography}{S9}
	
	\bibitem[S1]{Brandt2} E. H. Brandt, {Phys. Rev. B}
	\textbf{60}, 11939 (1999).
	
	\bibitem[S2]{Putzke2} C. Putzke, P. Walmsley, J.D. Fletcher, L. Malone, D. Vignolles, C. Proust, S. Badoux, P. See, H. E. Beere,
	D. A. Ritchie, S. Kasahara, Y. Mizukami, T. Shibauchi, Y. Matsuda, and A. Carrington, {Nat. Commun.} \textbf{5}, 5679 (2014).
	
	\bibitem[S3]{Yamashita2}
	{T.}~{Yamashita}, {T.}~{Takenaka}, {Y.}~{Tokiwa}, {J.~A.}~{Wilcox}, {Y.}~{Mizukami}, {D.}~{Terazawa}, {Y.}~{Kasahara}, {S.}~{Kittaka}, {T.}~{Sakakibara}, {M.}~{Ko{\'n}czykowski}, {S.}~{Seiro}, {H.~S.}~{Jeevan}, {C.}~{Geibel}, {C.}~{Putzke}, {T.}~{Onishi}, {H.}~{Ikeda}, {A.}~{Carrington}, {T.}~{Shibauchi},\ and\ {Y.}~{Matsuda}, Sci. Adv. \textbf{3}, e1601667 (2017).
	
	
\end{thebibliography}
\end{document}